\documentclass[structabstract]{aa}  
\usepackage{graphicx}
\usepackage{txfonts}
\usepackage{wrapfig, subfig, threeparttable, rotating}
\usepackage{natbib}
\bibpunct{(}{)}{;}{a}{}{,} %to follow the A&A style 
\begin{document}
   \title{The {\it Herschel}\thanks{{\it Herschel} is an ESA space observatory with science instruments provided by European-led Principal Investigator consortia and with important participation from NASA.} HIFI water line survey in the low-mass proto-stellar outflow L1448}

   \author{G. Santangelo
          \inst{1}
          \and
          B. Nisini
          \inst{1}
          \and
          T. Giannini
          \inst{1}
          \and
          S. Antoniucci
          \inst{1}
          \and
          M. Vasta
          \inst{2}
          \and
          C. Codella
          \inst{2}
          \and
          A. Lorenzani
          \inst{2}
          \and
          M. Tafalla
          \inst{3}
          \and
          R. Liseau
          \inst{4}
          \and
          E.F. van Dishoeck
          \inst{5,6}
          \and
          L.E. Kristensen
          \inst{5}
          }

   \institute{Osservatorio Astronomico di Roma, via di Frascati 33, 
              00040 Monteporzio Catone, Italy\\
              \email{gina.santangelo@oa-roma.inaf.it}
         \and
              Osservatorio Astrofisico di Arcetri, Largo Enrico Fermi 5, I-50125 Florence, Italy
         \and
              Observatorio Astron\'omico Nacional (IGN), Calle Alfonso XII 3, E-28014 Madrid, Spain
         \and
              Department of Earth and Space Sciences, Chalmers University of Technology, Onsala Space Observatory, 439 92 Onsala, Sweden
         \and
              Leiden Observatory, Leiden University, PO Box 9513, 2300 RA Leiden, The Netherlands
         \and
              Max Planck Institut f{\"u}r Extraterrestrische Physik, Giessenbachstrasse 1, 85748 Garching, Germany
             }

   \date{Received ; accepted } %month n.o, year

% \abstract{}{}{}{}{} 
% 5 {} token are mandatory
 
  \abstract
  % context heading (optional)
  % {} leave it empty if necessary  
   {}
  % aims heading (mandatory)
   {
      As part of the WISH (Water In Star-forming regions with {\it Herschel}) key project, systematic observations of 
H$_2$O transitions in young outflows are being carried out, with the aim of understanding 
the role of water in shock chemistry and its physical and dynamical properties.

In this paper, we report on the observations of several ortho- and para-H$_2$O lines 
performed with the HIFI instrument towards 
two bright shock spots (R4 and B2) along the outflow driven by the L1448 low-mass proto-stellar system, located in the Perseus cloud. These data are used 
to identify the physical conditions giving rise to the H$_2$O emission and infer any dependence with velocity. 
   }
  % methods heading (mandatory)
   {We used a large velocity gradient (LVG) analysis to derive the main physical 
parameters of the emitting regions, namely $n$(H$_2$), $T_{\rm kin}$, $N$(H$_2$O) and emitting-region size.
Comparison has been made with other main shock tracers, such as CO, SiO and H$_2$
and with shock models available in the literature.
}
  % results heading (mandatory)
   {These observations provide evidence that the observed water lines probe a 
warm ($T_{\rm kin} \sim$~400-600 K) and very dense ($n\sim$10$^6-10^7$ cm$^{-3}$) gas,
not traced by other molecules, such as low-$J$ CO and SiO, but rather traced by 
mid-IR H$_2$ emission.
In particular, H$_2$O shows strong differences with SiO in the excitation conditions and in the 
line profiles in the two observed shocked positions, pointing to chemical variations 
across the various velocity regimes and chemical evolution in the different shock spots.
Physical and kinematical differences can be seen at the two shocked positions.
At the R4 position, two velocity components with different excitation can be distinguished, with the 
component at higher velocity (R4-HV) being less extended and less dense than the 
low velocity component (R4-LV). 
H$_2$O column densities of about 2~10$^{13}$ and 4~10$^{14}$~cm$^{-2}$ have been derived for the 
R4-LV and the R4-HV components, respectively. 
The conditions inferred for the B2 position are similar to those of the R4-HV component,
with H$_2$O column density in the range 10$^{14}- 5$~$10^{14}$~cm$^{-2}$, corresponding to H$_2$O/H$_2$ 
abundances in the range $0.5-1$~10$^{-5}$.
The observed line ratios and the derived physical conditions seem to be more consistent with excitation in a low 
velocity J-type shock with large compression rather than in a stationary C-shock, although none 
of these stationary models seems able to reproduce all the characteristics of the observed emission.
   }
  % conclusions heading (optional), leave it empty if necessary 
   {}


\authorrunning{Santangelo et al.}
\titlerunning{The {\it Herschel} HIFI water line survey in L1448}
   \maketitle
%
%________________________________________________________________

\section{Introduction}
\label{sect:intro}

Strong radiative interstellar shocks are produced by the interaction 
of supersonic mass ejections from young stellar objects with the dense
ambient cloud. The observational signature of these shocks are 
the bright line emissions from molecules and atoms, whose
excitation conditions and relative abundances reveal fundamental
information on the type of interaction, and on the physical properties
of both the jets and the ambient medium. In the dense environments
of young proto-stars (the so-called Class~0 sources) the 
gas cooling occurs through emission of
H$_2$, CO and H$_2$O over a wide range of wavelengths, spanning from near-IR
to sub-mm wavelengths \citep[e.g.][]{kaufman1996,flower2010}.
Among these main coolants of outflow 
shocks, water is the most sensitive to local
conditions, since its abundance can vary by order of magnitudes 
through the shock lifetime \citep[e.g.][]{bergin1998}. Due to the difficulty of
observing water with ground-based facilities, the study of water is restricted to
IR/sub-mm telescopes from space. Dedicated sub-mm satellites,
such as SWAS and Odin, have allowed, for the first time, to observe 
the ground state ortho-H$_2$O at 557~GHz and to compare
its profile and abundance with that of CO \citep[e.g.][]{franklin2008,bjerkeli2009}. 
The poor spatial resolution of these facilities, together with the
restriction of observing a single line, have however prevented the study 
of the water excitation conditions and their variations along outflows. 
Through the Infrared Space Observatory (ISO),
on the other hand, it has been possible to infer the physical conditions
of the warm component giving rise to strong water emission \citep[e.g.][]{liseau1996,giannini2001}.
Nonetheless, the limited spatial and spectral resolution have not allowed
the association of this warm gas with a specific kinematical component.
The {\it Herschel Space Observatory} 
is now able to overcome these limitations, thanks to its improved spatial
and spectral resolution over a large wavelength range. 

As part of the WISH \citep[Water In Star-forming regions with {\it Herschel},][]{vandishoeck2011} key program, 
we are undertaking systematic observations
of H$_2$O transitions in young outflows, employing the two
spectrometers on board {\it Herschel}, HIFI and PACS. One of our main targets
of investigation is the L1448-mm outflow. This flow, powered by
a low-luminosity (11~$L_{\odot}$) Class~0 proto-star located in Perseus \citep[$d=235$~pc,][]{hirota2011}, is known to be a
strong water emitter on the basis of ISO observations \citep{nisini1999,nisini2000}.
HIFI observations of water in the central region of the L1448 outflow 
(within 20$^{\prime\prime}$ radius from the source)
have been presented by \cite{kristensen2011}, who focused on the H$_2$O properties in
the extreme high velocity gas associated with the source collimated jet.

In this paper, we present HIFI observations of several H$_2$O
transitions obtained in two active shocked regions along the L1448 flow.
Our aim is to study the excitation conditions of H$_2$O in the shocked gas, 
exploring, in particular, variations of excitation with velocity. 
We want in this way to study the physical and chemical conditions of the interstellar 
medium, which has been 
affected by the shock.

%
%________________________________________________________________

\section{Observations and data reduction}
\label{sect:obs}

   \begin{figure}%[ht]
   \centering
   \includegraphics[angle=-90,width=0.45\textwidth]{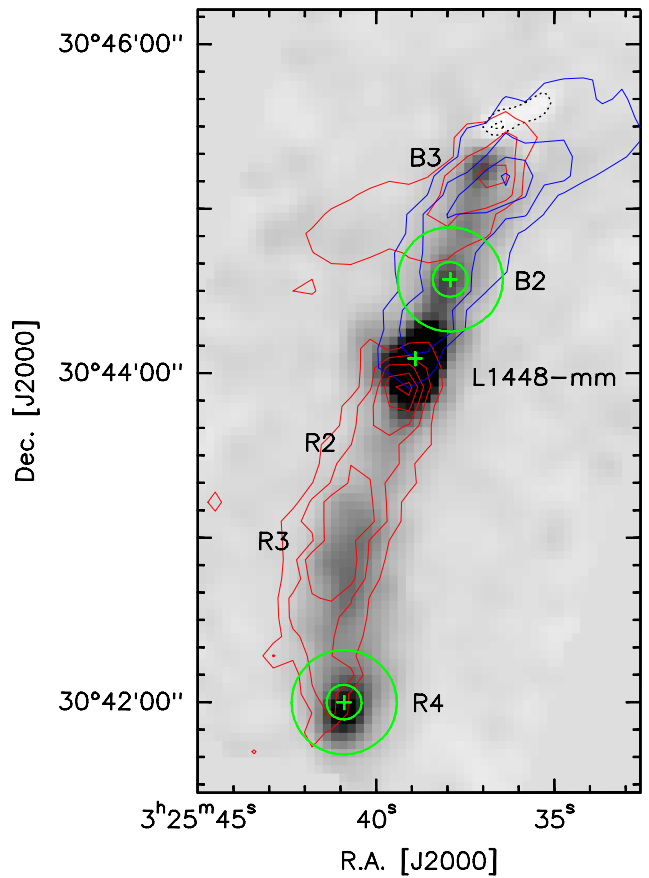}
   \caption{PACS image of L1448 at 1670 GHz
in false colors, with the negative contours in dotted line, and 
the JCMT CO(3-2) emission (Nisini et al. 2011, in prep.) in contours: the blue-shifted emission is integrated 
between -100 and 4~km~s$^{-1}$ and the red-shifted emission between 6 and 100~km~s$^{-1}$.
The central position of the map, as well as the positions chosen for the HIFI line survey (R4 and B2), 
are marked with a green cross.
For the latter two positions the largest and the smallest HIFI beam sizes are
shown in green.}
         \label{fig:outflow}
   \end{figure}

%\begin{sidewaystable*}
\begin{table*}
\caption{Summary of the HIFI line survey.} 
\label{table:h2o_summary}
\centering
\begin{threeparttable}
\renewcommand{\footnoterule}{}
\begin{tabular}{l@{ }r c r r@{.}l c c | r r c c | r r c } 
\hline\hline
& & & && & & & \multicolumn{2}{c}{R4} & R4-LV\tnote{a} & R4-HV\tnote{a} & \multicolumn{3}{c}{B2\tnote{a}} \\
\multicolumn{2}{c}{Line} & Band & $\nu$ & \multicolumn{2}{c}{E$_{\rm u}/k_B$} & HPBW & $\eta_{\rm mb}$ & $t_{\rm obs}$\tnote{b} & rms\tnote{c} & $\int T_{\rm mb}$ d$v$ & $\int T_{\rm mb}$ d$v$ & $t_{\rm obs}$\tnote{b} & rms\tnote{c} & $\int T_{\rm mb}$ d$v$ \\
& & & (GHz) & \multicolumn{2}{c}{(K)} & (arcsec) & & (min) & (mK) & (K km s$^{-1}$) & (K km s$^{-1}$) & (min) & (mK) & (K km s$^{-1}$)\\
\hline
o-H$_2$O     & 1$_{10}-1_{01}$  & 1 &   556.9 &  26&7  & 38.1 & 0.75 &  8.2  &  9 &  4.79$\pm$0.04  &  6.31$\pm$0.07  &  8.5 &  9  &  9.59$\pm$0.09  \\ 
             & 2$_{12}-1_{01}$  & 6 &  1669.9 &  80&1  & 12.7 & 0.71 & 21.2  & 56 &  4.56$\pm$0.27  &  9.90$\pm$0.37  & 20.3 & 100 & 6.30$\pm$1.00   \\
             & 3$_{12}-3_{03}$  & 4 &  1097.4 & 215&1  & 19.3 & 0.74 & 15.6  & 16 &  1.23$\pm$0.08  &  0.68$\pm$0.11  & 15.5 & 30  & $<$4.5\tnote{d} \\
p-H$_2$O     & 1$_{11}-0_{00}$  & 4 &  1113.3 &  53&4  & 19.0 & 0.74 & 15.6  & 19 &  3.62$\pm$0.08  &  5.34$\pm$0.12   & 15.5 & 26 &  4.85$\pm$0.27  \\ 
             & 2$_{11}-2_{02}$  & 2 &   752.0 & 136&9  & 28.2 & 0.75 &  8.4  & 16 &  2.45$\pm$0.07  &  1.32$\pm$0.11   &  8.6 & 19 &  1.76$\pm$0.19  \\
o-H$_2^{18}$O & 1$_{10}-1_{01}$  & 1 &   547.7 &  26&2  & 38.7 & 0.75 & 57.7  &  3 & $<$0.18\tnote{d} & $<$0.36\tnote{d} &     &    &                 \\
NH$_3$       & 1$_{0}-0_{0}$   & 1  &  572.5  &  27&5  & 37.0 & 0.75 &  8.2 &   9 & $<$0.54\tnote{d} & $<$1.08\tnote{d} & 8.5 &  9 & 0.47       \\
\hline
\end{tabular}
\begin{tablenotes}
\item[a] Integrated intensities of the lines in the two velocity components at R4 
(between 0 and 20~km~s$^{-1}$ for R4-LV and between 20 and 60~km~s$^{-1}$ for R4-HV) 
and total integrated intensities of the lines at B2. 
\item[b] Total observing time: on $+$ off $+$ overheads.
\item[c] All spectra were smoothed to a velocity resolution of 1~km~s$^{-1}$.
\item[d] Upper limits to the integrated intensities, computed from the 3$\sigma$ values of the spectra and the 
line widths of the respective velocity components.  
\end{tablenotes}
\end{threeparttable}
\end{table*}
%\end{sidewaystable*}

Figure~\ref{fig:outflow} shows the JCMT CO(3-2) emission in contours, overlaid on 
the PACS map of the H$_2$O 2$_{12}-1_{01}$ emission at 1670~GHz towards L1448 (Nisini et al. 2011, in prep.).
The H$_2$O map exhibits several emission peaks, which correspond to  
positions of active shock regions, named B2-B3 (for the blue-shifted lobe) and R2-R3-R4
(for the red-shifted lobe), following the nomenclature of \cite{bachiller1990}, based on the analysis of CO line profiles.
For our water line survey we selected the shock spots B2 and R4: the first corresponds
to a bright H$_2$ emission region \citep{davis1995}, arising from the interaction
of the high velocity jet from the mm source with the ambient medium; while the second
is located at the end of the red-shifted lobe and is spatially associated with
the bow-shock seen in SiO by \cite{dutrey1997}.
Their offsets with respect to the central driving source L1448-mm 
($\alpha$(J2000)=03$^{\rm h}$25$^{\rm m}$38$.\!\!^{s}$9, $\delta$(J2000)=+30$^\circ$44$^{\prime}$05$.\!\!^{\prime\prime}$4)
are respectively ($-16^{\prime\prime}$, 34$^{\prime\prime}$) for B2 and (26$^{\prime\prime}$, $-128^{\prime\prime}$) for R4.

A survey of several ortho- and para-H$_2$O lines has been conducted with the HIFI heterodyne instrument 
\citep{degraauw2010} on board the {\it Herschel} Space Observatory \citep{pilbratt2010},
towards these two positions. 
The survey comprises lines with excitation energies 
ranging from 27 to 215~K and also includes the ortho-H$_2^{18}$O 1$_{10}-1_{01}$ transition
at 547.7~GHz, useful to infer opacity effects.

The data were processed with the ESA-supported package HIPE\footnote{HIPE is a joint development 
by the {\it Herschel} Science Ground Segment Consortium, consisting of ESA, the NASA {\it Herschel} 
Science Center, and the HIFI, PACS and SPIRE consortia.} 
\citep[{\it Herschel} Interactive Processing Environment,][]{ott2010} for calibration. 
The calibration uncertainty is taken to be 20\%.
Severe baseline problems have been found at the B2 position in Band 6 and Band 4, 
corresponding to the ortho-H$_2$O 2$_{12}-1_{01}$ line at 1670~GHz and
the 3$_{12}-3_{03}$ line at 1097~GHz, respectively.
We obtained, using the experimental 
``matching technique''\footnote{The techniques are being developed at the HIFI ICC; 
see http://herschel.esac.esa.int/twiki/bin/view/Public/HifiCalibration\\
Web?template=viewprint for references.
} 
for electronic standing waves, 
a tentative detection for the former line at a $\sim$3$\sigma$ level, 
which is consistent with the PACS map at 1670~GHz, presented in Fig.~\ref{fig:outflow};
for the latter instead only an upper limit has been derived.

Further reduction of all the spectra, including baseline subtraction, and 
the analysis of the data were 
performed using the GILDAS\footnote{http://www.iram.fr/IRAMFR/GILDAS/} software.
H- and V-polarizations were co-added after inspection; no significant differences were found 
between the two data sets.
The calibrated $T^*_A$ scale from the telescope was converted into the $T_{\rm mb}$ scale using the 
main-beam efficiency factors provided by Roelfsema et al. (2011, submitted)\footnote{see also
http://herschel.esac.esa.int/twiki/bin/view/Public/\\
HifiCalibrationWeb?template=viewprint}
and reported in Table~\ref{table:h2o_summary}.
The beam sizes range from $\sim$13$^{\prime\prime}$ to 39$^{\prime\prime}$.
The largest and the smallest beam sizes of the HIFI observations are marked
as green circles in Fig.~\ref{fig:outflow}, for each observed position.
At the velocity resolution of 1~km~s$^{-1}$, the rms noise range is 3--100~mK ($T_{\rm mb}$ scale).

   \begin{figure*}%[h!]
   \centering
   \includegraphics[width=0.7\textwidth]{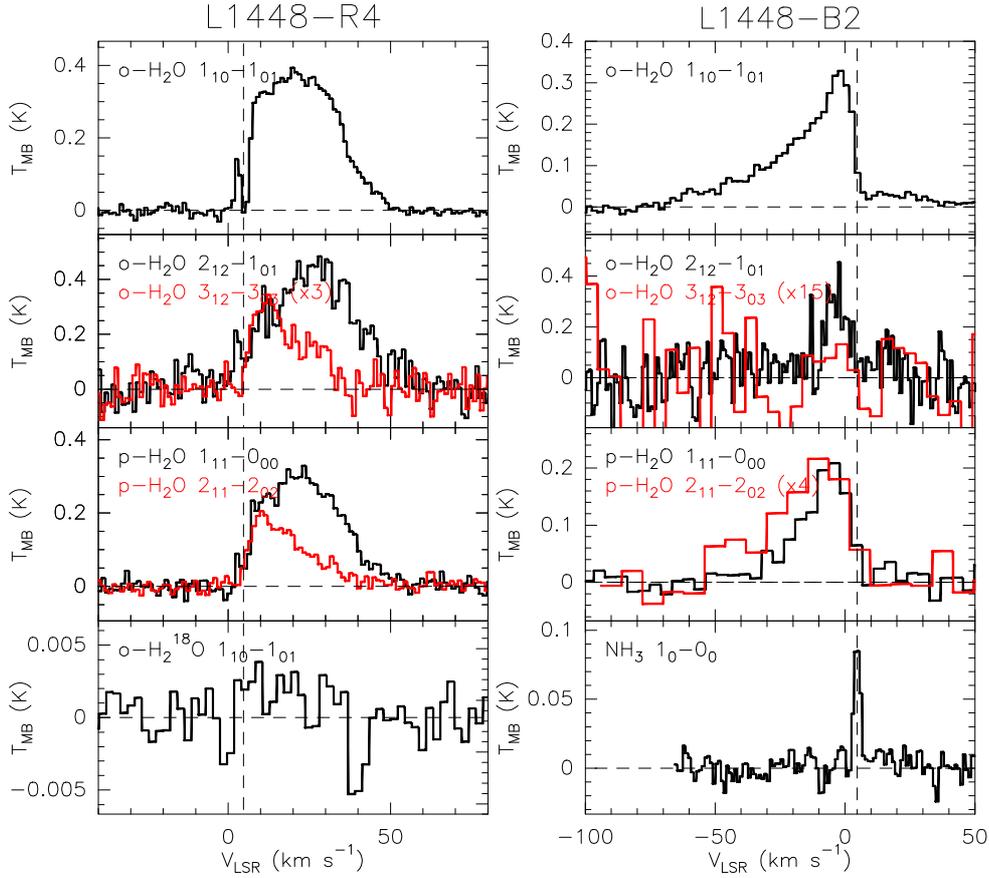}
   \caption{HIFI spectra at the R4 (left) and B2 (right) positions. Note the different profiles
for low-lying (black) and high-lying (red) levels at the R4 position.
}
   \label{fig:h2o_spettri}
   \end{figure*}

Transitions from other molecules have also been observed in the HIFI bands centered
on the water lines, namely NH$_3$(1$_0-0_0$), $^{13}$CO($10-9$), HCO$^+$($6-5$) and C$^{18}$O($5-4$).
The only detected line among these is the NH$_3$ transition at the B2 position, 
while for all the other transitions we have non-detections
at a rms noise level (at the velocity resolution of 1~km~s$^{-1}$) of 4~mK (HCO$^+$ $6-5$ and C$^{18}$O $5-4$)
and $20-26$~mK ($^{13}$CO $10-9$).
A summary of the observations, including frequencies ($\nu$), beam sizes (HPBW), main-beam 
efficiencies ($\eta_{\rm mb}$) and velocity-integrated intensities is given in 
Table~\ref{table:h2o_summary}.

%
%______________________________________________________________

\section{Line profiles}
\label{sect:line_profiles}

The spectra of the H$_2$O lines observed at the R4 and B2 spots are presented in 
Fig.~\ref{fig:h2o_spettri}, respectively on the left and right panels.

\begin{figure}%[ht]
   \centering
   \includegraphics[angle=-90,width=0.45\textwidth]{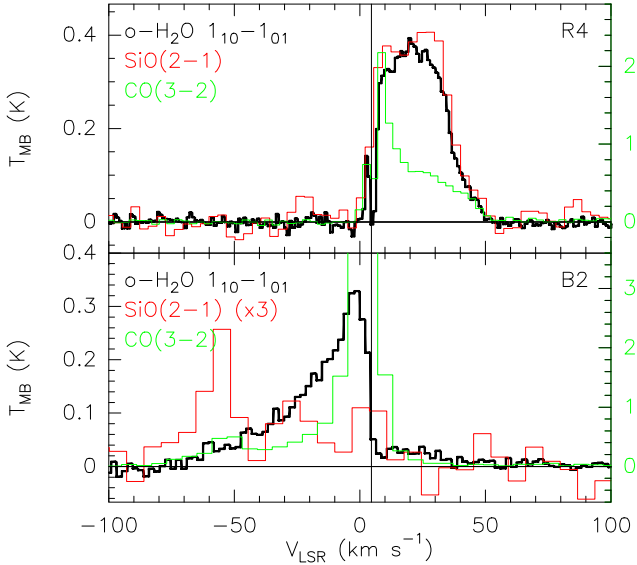}
   \caption{Overlay between the HIFI H$_2$O 557~GHZ line (black), IRAM-30m SiO(2-1) 
line (red) from \cite{nisini2007} and JCMT CO(3-2) line (green) from Nisini et al. (2011, in prep.),
towards the R4 ({\it upper}) and the B2 ({\it lower}) positions.
}
   \label{fig:h2o_sio_co}
   \end{figure}

At the R4 position, the water line profiles reveal a broad emission extending up to
about 50~km~s$^{-1}$ with respect to the systemic velocity.
The overlay between water line profiles with different upper level energies shows
that the lines having $E_{\rm u}<$~137~K peak at velocities around 25~km s$^{-1}$,
while the lines with higher upper level energies peak at lower velocities (around 10~km~s$^{-1}$).
The different profiles of lines at different excitation could in principle be caused
by large self-absorptions in the low velocity range of those lines connecting with the
ground state. This could be due to the presence of large columns of cold water
along the line of sight. However, we do not think this is the case, since the effect
is not observed in the spectra obtained at the B2 position. In addition, as we will show 
later in Fig.~\ref{fig:h2o_sio_co},
the profile of the H$_2$O 1$_{10}-1_{01}$ line at 557~GHz matches well the SiO $J$=2-1 line profile, which does not
show any evidence of self-absorption even at the systemic velocity.
The two velocity regimes at R4 (and correspondingly the different excitation lines) are thus probably
tracing gas with different physical conditions.

In particular, from the comparison of the shape of the profiles shown in Fig.~\ref{fig:h2o_spettri},
it can be noted that the lines at higher excitation (namely the o-H$_2$O at 1097~GHz and the p-H$_2$O at 752~GHz)
have a simple triangular profile peaking at low velocity, similar to those observed also in B2, while the other lines
present an additional component peaking at higher velocity, overlaid on the low velocity triangular profile. 
It thus seems that the observed profiles results from the superposition of two gas components, having different
kinematical and excitation properties, seen along the line of sight.

Finally, 
the non-detection of the ortho-H$_2^{18}$O 1$_{10}-1_{01}$ line at 548 GHz is shown in the 
bottom-left panel of Fig.~\ref{fig:h2o_spettri}, 
with a rms noise of about 3 mK ($T_{\rm mb}$ scale) at 1~km s$^{-1}$ velocity resolution. 
This gives a constraint on the H$_2$O 1$_{10}-1_{01}$ optical depth. 
Assuming an oxygen isotope ratio $^{16}$O/$^{18}$O$\sim 560$ from \citet[for local interstellar medium]{wilson1994}
and under the assumption that the two lines have the same $T_{\rm ex}$,
we find that $\tau_{\rm 557GHz} <$~15 at the peak of the H$_2$O emission
(based on H$_2^{18}$O 3$\sigma$ upper limit). 

In contrast with R4, all the detected lines at the B2 position
show a similar triangular line profile with a line wing extending up to about $-50$~km~s$^{-1}$.
It is interesting to point out that, among the detected H$_2$O lines, the highest excitation line 
(2$_{11}$-2$_{02}$ with $E_{\rm u}=137$~K) seems to show 
emission extending at high velocity, around $-45$~km~s$^{-1}$, at a level of $\lesssim 3$~$\sigma$.
This may indicate the presence of high excitation gas at high velocity.

The bottom-right panel of Fig.~\ref{fig:h2o_spettri} also shows the NH$_3$(1$_0-0_0$) emission line
detected at B2. Ammonia emission is confined at the systemic velocity of about 5~km~s$^{-1}$,
testifying to its origin from the
molecular cloud and not from the outflow \citep[see also][]{bachiller1990}.

Figure~\ref{fig:h2o_sio_co} shows the comparison between the H$_2$O 1$_{10}-1_{01}$ line profile in black
and that of SiO $v$=0, $J$=2-1 in red and the CO(3-2) in green 
\citep[2011]{nisini2007}, towards both investigated positions.
The figure highlights the very different line profile behaviour at the two positions.
In particular, the profile of the H$_2$O line at 557~GHz follows quite closely
the SiO profile at the R4 position, while at the B2 position the SiO emission
is enhanced at the extreme high velocities (EHVs, $v \sim$~55~km~s$^{-1}$).
Conversely, there is no water enhancement above the wing profile at the highest velocities, which suggests that 
the water emission from the EHV gas is observed only closer to the driving source (see also
\citealt{kristensen2011}; Nisini et al. 2011, in prep.).
A discussion of the different behaviour of H$_2$O and SiO line profiles at the two observed positions is given 
later, in Sect.~\ref{sect:discussion}.

At both positions, the comparison with the CO(3-2) emission points out that H$_2$O always has excess 
emission with respect to CO at intermediate velocities, suggesting that H$_2$O and CO could trace 
different gas components. This will be further discussed in Sect.~\ref{sect:discussion}.

%
%______________________________________________________________

\section{Excitation analysis}
\label{sect:analysis}

The physical conditions of the observed emitting regions have been
investigated by comparing the observed intensities with predictions from
the RADEX escape probability code in plane parallel geometry \citep{vandertak2007}. 
Although the plane parallel geometry is not able to reproduce the line profiles \cite[e.g.][]{bjerkeli2011}, 
it can be used as a first approximation for this study.
The molecular data were taken from the Leiden 
Atomic and Molecular Database \citep[LAMBDA\footnote{http://www.strw.leidenuniv.nl/~moldata/},][]{schoier2005},
with the collisional rate coefficients from \cite{faure2007}.

As pointed out in Sect.~\ref{sect:line_profiles}, the profiles at the R4 position present
two distinct components showing a clear change in excitation with velocity.
We therefore need to separate the two velocity components and analyze them separately.
An approach to separate the two velocity components in each line may be 
to fit the triangular component, peaking at low velocity, in each water line and subtract it from the rest of the line.
However, due to the fact that fitting a triangular shape to the water lines is arbitrary, we opted for a simpler approach. 
We divided the emission at R4 into two velocity components: 
a low--velocity component (R4-LV hereafter, between 0 and 20 km~s$^{-1}$) and a high--velocity one 
(R4-HV hereafter, between 20 and 60 km~s$^{-1}$).
As described in the Appendix, the results of the analysis will be little affected by the method we 
used to separate the two velocity contributions, due to the fact that both components dominate 
the relative velocity range we considered. 
 
For B2, instead, due to the low signal-to-noise and the non-detection of one H$_2$O line out of five, 
we had too little information to perform an excitation analysis as a function of velocity,
and therefore we have integrated the emission over the whole velocity range.
The observed velocity-integrated intensities for each component are given in Table~\ref{table:h2o_summary}.

We have compared the integrated intensities with a grid of RADEX models,
constructed by varying the parameters in the following ranges: 
$n$(H$_2)= 10^4 - 10^8$~cm$^{-3}$, $N$(ortho-H$_2$O)$= 10^{12} - 10^{16}$~cm$^{-2}$, 
$T_{\rm kin}$=100--1600~K and ortho to para ratio $o/p=3, 2.5, 2, 1.5$. 
Line widths (FWZI) of 20, 40 and 50~km~s$^{-1}$ have been adopted for R4-LV, R4-HV and B2,
respectively.
Given the different beam sizes of the observations, we have also considered 
the size of the emission region ($\Theta$) as an additional parameter: 
we varied it from almost point-like (few arcsec) to extended emission.
A Gaussian emitting region, as well as a Gaussian beam, were assumed to 
correct the emitting size for beam dilution effects.

\subsection{R4 position}
\label{subsect:analysis_R4}

The comparison of the observations with the RADEX grid of models
shows that only a limited range of physical parameters is able to 
simultaneously reproduce both the line ratios and the 
line intensities in each component (see Figs.~\ref{Afig:LV-diagnostic}
and \ref{Afig:HV-diagnostic} in the Appendix).
In Fig.~\ref{fig:mod-obs} these best fit models are compared with the measured integrated intensities. 
For the R4-LV component, we are able to reproduce the observations,
within the calibration errors, assuming only extended 
warm gas ($T_{\rm kin} \sim 600$~K) having a very high density $n$(H$_2) \sim 10^7$~cm$^{-3}$ 
(model LV-1 in Table~\ref{table:h2o_best_fit}). The implied column density
is of the order of $N$(H$_2$O$) \sim 2$~$10^{13}$~cm$^{-2}$, which indicates that the observed
lines have low opacities ($\tau_{\rm 557GHz} <$~0.1), consistent with the non-detection of the H$_2^{18}$O 
transition. 

For the R4-HV component, on the other hand, there is more degeneracy in the physical
parameters (see Fig.~\ref{Afig:HV-diagnostic}) and the observations can be reproduced either by 10$^7$~cm$^{-3}$ gas
at low temperature ($T_{\rm kin} \sim 150$~K, model HV-1) or by gas as warm as in the R4-LV
component, but having a lower density of $\sim 10^6$~cm$^{-3}$ (model HV-2).
It is therefore not possible, at this stage, to infer whether the lower excitation
of the R4-HV component with respect to the R4-LV component, evidenced by the comparison
of the line profiles, is due to a temperature or a density effect. In both cases,
the R4-HV component appears to be more compact ($\sim 13-21^{\prime\prime}$) and with a higher 
column density $N$(H$_2$O$)=$~($7-40$)~$10^{13}$~cm$^{-2}$, 
with respect to the R4-LV gas. 

The different sizes found in the R4-LV and R4-HV components are consistent
with the PACS map at 1670~GHz presented in Fig.~\ref{fig:outflow}.
The map in fact shows that the 1670~GHz emission at R4 is dominated by a 
compact emission,
which may correspond to R4-HV, superimposed on a more 
extended weak emission, which may correspond to R4-LV.

The physical conditions derived for the two components, in particular the
very high density, can be compared with those inferred from other shock tracers,
to see whether the observed H$_2$O emission probes the same gas. 
Given the observed similarity between the H$_2$O 557 GHz and the SiO(2-1) line profiles
shown in Fig.~\ref{fig:h2o_sio_co}, one would be tempted to conclude that
the two species have similar excitation conditions. 
To test this hypothesis, we have used the SiO multi-line observations
at R4 presented by \cite{nisini2007} to derive
temperature and H$_2$ density of the SiO gas in
both the R4-LV and R4-HV components. We have used a grid of LVG models, 
constructed from the RADEX code, and assumed an emission size of the order of 
15$^{\prime\prime}$, for both components, estimated from the IRAM Plateau de Bure (PdB)
channel maps of the SiO(2-1) emission, presented by \cite{dutrey1997}.

\begin{table}[ht]
\caption{Summary of the models derived for each component.}
\label{table:h2o_best_fit}
\centering               
\begin{threeparttable}
\renewcommand{\footnoterule}{}
\begin{tabular}{l l l c r r c} 
\hline\hline
Comp. & Model & $o/p$ & $T_{\rm kin}$ & $n$(H$_2)$ & $N$(H$_2$O)\tnote{a}  & $\Theta$ \\
& & & (K) & (cm$^{-3}$) & (cm$^{-2}$) & (arcsec) \\
\hline
R4-LV & {\bf LV-1}\tnote{b} & {\bf 3}   & {\bf 600} &   {\bf 10$^{7}$} & {\bf 2 10$^{13}$} & {\bf 37}  \\
      & LVsio-2\tnote{c} & 3   & 500 & 3 10$^{4}$ & 3 $10^{15}$      & 15   \\
\hline
R4-HV & HV-1 & 3   & 150 & 3 10$^{7}$ & 7 10$^{13}$      & 21   \\
      & {\bf HV-2}\tnote{b} & {\bf 3}   & {\bf 650} &   {\bf 10$^{6}$} & {\bf 4 10$^{14}$} & {\bf 13}  \\
      & HVsio-3\tnote{c} & 3   & 250 & 5 10$^{4}$ & 9 10$^{15}$      & 13   \\
      & HVsio-4\tnote{c} & 3   & 600 &   10$^{4}$ & 3 10$^{16}$      & 13   \\
\hline
B2    & B2-1  & 3   & 450 & 6 10$^{6}$ & 10$^{14}$      & 33   \\
      & {\bf B2-2}\tnote{b,d} & {\bf 3}   & {\bf 450} &   {\bf 10$^{6}$} & {\bf 5 10$^{14}$} & {\bf 17}  \\
\hline
\end{tabular}
\begin{tablenotes}
\item[a] Ortho- $+$ para-H$_2$O.
\item[b] The bold font highlights %the best model for each component
for each component the model that better fits all the observational constraints.
\item[c] Physical parameters ($T_{\rm kin}$, $n_{\rm H_2}$,$\Theta$) derived from the SiO analysis
(see Sect.~\ref{subsect:analysis_R4}).
\item[d] See Sect.~\ref{subsect:analysis_B2}.
\end{tablenotes}
\end{threeparttable}
\end{table}

   \begin{figure}[ht]
   \centering
   \includegraphics[angle=-90,width=0.48\textwidth]{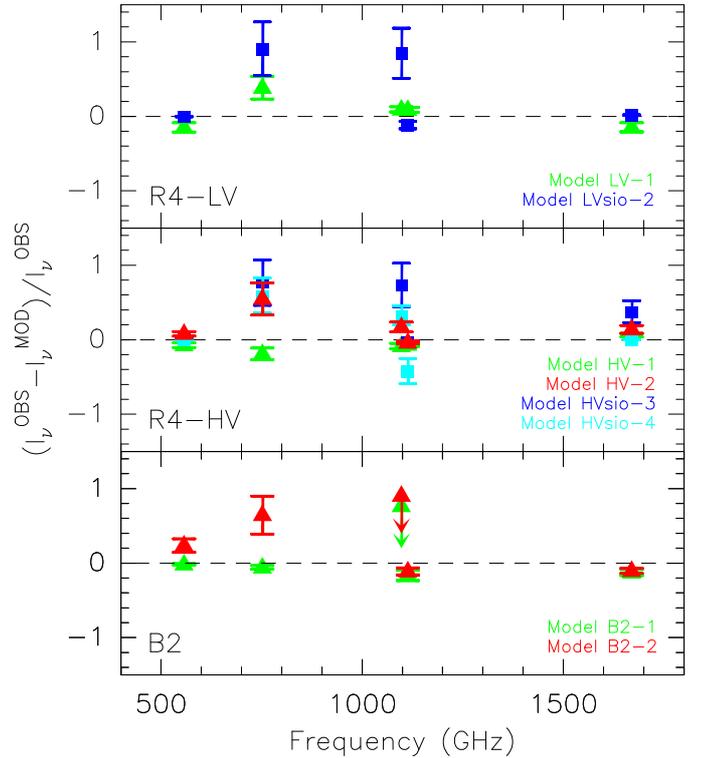} 
   \caption{{\it Top and middle panels}: Comparison between the observed water intensities 
   and the different models in Table~\ref{table:h2o_best_fit}, for R4-LV ({\it top}),
   R4-HV ({\it middle}) and B2 ({\it bottom}). 
The intensities predicted by the models have been corrected for the relative predicted filling factors. 
The given errors include 20\% of calibration error.
The triangles represent the models derived from H$_2$O and the squares from the SiO.
{\it Bottom panel}: Same comparison as the {\it top panels}, but for B2. 
The upper limit of the non-detected H$_2$O line at 1097~GHz is also shown for each model.
}
   \label{fig:mod-obs}
   \end{figure}

The best fit models that reproduce the SiO emission in the two velocity components 
are given in Table~\ref{table:h2o_best_fit} (LVsio-2 for R4-LV and HVsio-3 and HVsio-4 for R4-HV).
The models trace warm gas ($T_{\rm kin}$ between 250 and 600~K) but with H$_2$ densities of only ($1-5$)~10$^{4}$~cm$^{-3}$, 
thus two order of magnitudes lower than those derived from the H$_2$O analysis.
In particular, the H$_2$ densities are lower than 
those derived from \cite{nisini2007}, $n$(H$_2) \approx 2.5$~10$^5$~cm$^{-3}$
with $T=200$~K, and this is probably due to beam-filling effects that were 
not taken into account in \cite{nisini2007}.
They assumed in fact a beam-filling factor equal to 1, which implies that the 
line intensities are reproduced with a lower column density and 
consequently a higher particle density.

We have then investigated how well the conditions estimated from the SiO emission reproduce the water lines,
by varying the H$_2$O column density in order to match the 557~GHz H$_2$O line.
The results are visualized in Fig.~\ref{fig:mod-obs}. 
For R4-LV, the SiO best fit model underestimates the intensities of the water lines 
at higher excitation (para-H$_2$O 2$_{11}-2_{02}$ at 752~GHz 
and ortho-H$_2$O 3$_{12}-3_{03}$ at 1097~GHz). 
This could indicate that SiO is tracing an additional low-velocity gas component 
at lower excitation than that probed by H$_2$O. 
Finally, for R4-HV we considered two possible SiO models 
(HVsio-3 and HVsio-4 in Table~\ref{table:h2o_best_fit}) and both of them reproduce 
reasonably well only four out of the five H$_2$O lines.

The high density regime found by the best-fit models, based only on the water lines, are more
consistent with the conditions derived from {\it Spitzer} mid-IR H$_2$ observations
along the L1448 flow \citep{giannini2011}. Although the R4 position is not covered
by these observations, the other H$_2$ shocked spots in the red-shifted lobe 
are consistent with a warm gas at densities of the order of 10$^6$~cm$^{-3}$.

\subsection{B2 position}
\label{subsect:analysis_B2}

At the B2 position we detect four (the 1670~GHz water line is however only a tentative detection) 
out of the five targeted transitions:
therefore we have too little information to constrain all the parameters
of the fit (see Fig.~\ref{Afig:B2-diagnostic}
in the Appendix).
In order to overcome this problem, we assume that the water
emission originates from the same gas emitting the mid-IR H$_2$ lines, 
observed with Spitzer \citep{giannini2011}.
This assumption is supported by the results found in other 
outflows observed within the WISH program \citep[e.g.][Tafalla et al. 2011, in prep.]{nisini2010}.
The H$_2$ rotational transitions at low-$J$ indicate a temperature of about 450~K, 
while a non-LTE fit through all the H$_2$ lines suggests high densities, 
in excess of 10$^6$~cm$^{-3}$. 

   \begin{figure*}[ht]
   \centering
   \includegraphics[angle=-90,width=0.98\textwidth]{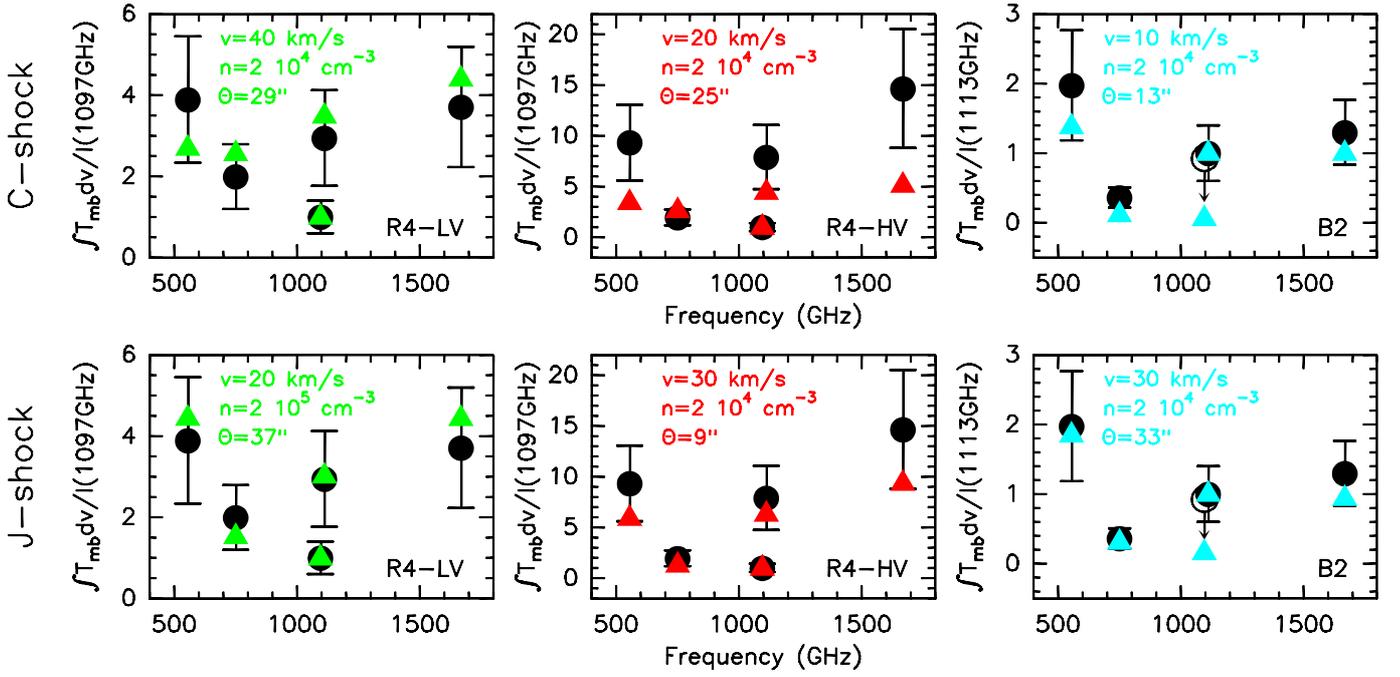}
   \caption{Comparison between the observed line ratios (circles) of the H$_2$O lines and 
the corresponding line ratios predicted (triangles) by the best-fit C-type shock ({\it top}) and the 
J-type shock ({\it bottom}) models.
For each type of shock (C- and J-), the model that best 
reproduces our observed line ratios is presented and the corresponding shock velocity, 
pre-shock density and emission size are marked.
The line ratios are normalized to the 1097~GHz line for R4-LV ({\it left}) and R4-HV 
({\it middle}) and to the 1113~GHz line for B2 ({\it right}). For B2, the detected lines 
are shown as black filled circles, while the black open circle represents the upper limit of the 
non-detected H$_2$O line at 1097~GHz (assuming a 3$\sigma$ upper limit and a line width of 
50~km~s$^{-1}$).
}
   \label{fig:mod-obs_shock}
   \end{figure*}

Keeping these constraints, 
the fit is still degenerate in the product $n$(H$_2$)$\times$$N$(H$_2$O), 
i.e. the data can be reproduced either by $n$(H$_2$)$\sim$6~$10^6$~cm$^{-3}$ and $N$(H$_2$O)$=10^{14}$~cm$^{-2}$
(model B2-1 in Table~\ref{table:h2o_best_fit} and Fig.~\ref{fig:mod-obs}) or 
$n$(H$_2$)$\sim$$10^6$~cm$^{-3}$ and $N$(H$_2$O)$=5$~$10^{14}$~cm$^{-2}$ (model B2-2).
The smaller size of the B2-2 model, with respect to the B2-1 model, seems
to be consistent with the rather compact PACS emission at the B2 position, as shown in Fig.~\ref{fig:outflow}.
Hence, the B2-2 model is indicated in Table~\ref{table:h2o_best_fit}
as the best model for the B2 component, even if it is associated with a worse fit. In fact, 
the comparison between the integrated intensities, predicted by the B2-2 model, as seen in 
Fig.\ref{fig:mod-obs}, shows that the intensities of the H$_2$O lines with the biggest beam sizes 
(at 557~GHz and 752~GHz) are slightly underestimated. This could be due to the fact
that the observations performed with these large beam sizes may collect
emission from additional components, especially from the source position, 
as can be evidenced from the PACS map of Fig.~\ref{fig:outflow}.

%
%______________________________________________________________

\section{Discussion}
\label{sect:discussion}

\subsection{Water abundance}
\label{subsect:abundance}

The presented excitation analysis shows that the observed HIFI water lines are
consistent with warm gas at high density having moderate H$_2$O column
densities that do not exceed $\sim$5~10$^{14}$ cm$^{-2}$. 
Similar water conditions have also been found in the L1157 flow \citep{vasta2011}.
These water column densities are significantly lower than those derived close to the 
L1448-mm source by \cite{kristensen2011}, who estimated 
$N$(H$_2$O) $\ga 10^{16}$~cm$^{-2}$ in all the
considered velocity components. 
This is in agreement with ISO observations of L1448 \citep{nisini1999,nisini2000}, 
where the strongest water emission is found towards the central source.
Moreover, this seems to suggest that the water emission from the EHV gas 
is only observed at high column densities, close to the driving source \citep[see][]{kristensen2011}, 
in contrast with other shock tracers.

H$_2$O/CO abundance ratios found by 
\cite{kristensen2011} are in the range 0.01--1, under the assumption that 
the two molecules are probing the same gas component.
Adopting the same assumption, we have applied the physical conditions derived
from water to the CO(3-2) data and computed the corresponding CO column 
densities using the RADEX code. The data were corrected for beam dilution effects and the same 
sizes as estimated from water were assumed. 
The derived $N$(CO) values are high in 
comparison with water, ranging between $\sim$ 5~10$^{16}$
at R4-LV and HV, and 2~10$^{17}$~cm$^{-2}$ at B2.
These values would imply a H$_2$O/CO abundance $\sim$10$^{-4}$--10$^{-2}$,
if one assumes the typical CO/H$_2$ ratio of 10$^{-4}$,
i.e. a water abundance extremely low for a shocked region. 
However the assumption that the H$_2$O 
and CO(3-2) emission come from the same region is likely not correct, 
given that they largely differ in both critical density (a factor of 10$^3$)
and excitation temperature, and given also their different line profiles 
(Fig.~\ref{fig:h2o_sio_co}). It is thus conceivable that CO(3-2)
traces mostly entrained ambient gas, while the H$_2$O emission comes
directly from the warm shocked gas.

A more direct estimate of the H$_2$O abundance of the gas component 
traced by the HIFI lines can be obtained comparing
the water column density with the H$_2$ column density of the 
same component, inferred from Spitzer observations \citep{giannini2011,dionatos2009}.
Only the B2 position was covered by these observations: the H$_2$ gas is 
stratified in temperature, with the gas at $T$$\sim$~400--500~K
having $N$(H$_2$) of the order of 5~10$^{19}$ cm$^{-2}$ in a $\sim$20$^{\prime\prime}$ beam.
The H$_2$O abundance at B2, relative to this gas component, can therefore be constrained to be 
$\sim0.5-1$~10$^{-5}$.
A similar abundance can be inferred for the R4-HV gas (whose estimated 
density is consistent with the average H$_2$ density given by \citealt{giannini2011}
along the flow), assuming that the same $N$(H$_2$) 
applies to this component.
This is clearly just a rough estimate, since we are comparing 
gas components at slightly different temperatures.
These abundances are lower than those expected in non-dissociative, 
stationary shocks, where most of the gas-phase oxygen is converted into water
\citep{kaufman1996}.
The derived H$_2$O abundances are however higher than interstellar values, 
indicating that the observed warm gas has been processed by shock-related chemical processes.

\subsection{Comparison with shock models}
\label{subsect:shock_models}

In order to better investigate the consistency of the observed emission with 
shock models, we have compared the observed intensity ratios with the grid of 
stationary C- and J-type shocks provided by \cite{flower2010}. This grid explores 
different shock velocities (from 10 to 40~km~s$^{-1}$) and two pre-shock 
densities (2~10$^{4}$ and 2~10$^{5}$~cm$^{-3}$).
For each type of shock (C- and J-), we have identified the model that better 
reproduces our observed line ratios. 
As in the case of the RADEX analysis, the emission size has been taken as an additional 
parameter, used to correct the intensity of lines with different beam sizes.
The results are shown in Fig.~\ref{fig:mod-obs_shock}, where
the observed and predicted line ratios are plotted with respect to the 1097~GHz
line for R4-LV and R4-HV and to the 1113~GHz line for B2.
From the figure one can see that the line ratios at R4-HV are
better reproduced by a J-type shock with a shock speed of 30~km~s$^{-1}$,
consistent with the line width of this velocity component.
The shock conditions of the R4-LV component are more degenerate:
in fact, both shock models seem to reproduce the data within the uncertainties. 
There are however different reasons to prefer the J-type shock model. 
First of all, the J-shock model has a shock speed of 20~km~s$^{-1}$ (versus 40~km~s$^{-1}$ for the 
C-shock), consistent with the
line width of the R4-LV component. In addition, the best-fit J-shock model have a
pre-shock density of 2 10$^5$~cm$^{-3}$, higher than that predicted
for R4-HV (2~10$^4$~cm$^{-3}$), and this is consistent with our LVG analysis,
showing that the difference in excitation between the R4-LV and R4-HV components
can be due to a different pre-shock density. The comparison between the post-shock density
derived from the LVG analysis and the pre-shock density predicted by the 
J-shock, indicates a compression factor $n_{\rm post}$/n$_{\rm pre} \sim$100, 
which is consistent with water emission originating from the post-shock gas 
regions further behind the shock front, where H$_2$, which has been initially 
dissociated, starts to reform, keeping the temperature to values around few 
hundreds of Kelvin \citep{flower2010}.

At the B2 position, only the J-type shock predicts a shock velocity consistent with 
the observed wide line profiles (i.e. 30~km~s$^{-1}$ against 10~km~s$^{-1}$ for the C-type shock), 
although both C- and J-type shock models reproduce the line ratios. 
This best-fit J-type model predicts a shock speed of 30~km~s$^{-1}$ and a pre-shock
density of 2 10$^4$~cm$^{-3}$. Several additional lines of evidence 
indicate that a dissociative shock must be present at this position. 
Spitzer-IRS observations
show that atomic emission from [\ion{Fe}{ii}]~26~$\mu$m and [\ion{S}{i}]~25~$\mu$m
is associated with this position \citep{neufeld2009,dionatos2009}. 
PACS observations reveal emission from OH and [\ion{O}{i}]~63$\mu$m (Santangelo et al.,
in prep.): this is a further indication that the gas-phase reactions 
that convert all the oxygen into water, activated at $T\gtrsim 300$~K,
have not been
completed and dissociative shocks must be present to have high OH emission \citep{neufeld1989}.

There are however also several inconsistencies between the considered models
and our observations: in particular, none of these models (either C- or J-type) is able  
to reproduce the absolute intensities of the observed lines. 
Indeed, the comparison between the predicted and observed
intensities would imply emission sizes much smaller (of the order of few arcsec) 
than those fitted through the line ratios.
We believe that these discrepancies are due to the simplified nature of the adopted shock models.
In fact, a single, plane-parallel shock model
is unlikely to provide a completely satisfactory representation of the line emission
from a complex shock structure, where both geometrical and temporal effects can significantly
alter the emergent line intensity and the final chemical abundance.
In particular, several studies have shown that shocks from
young outflows have not reached the steady state yet, and thus are
better represented with non-stationary shock conditions that combine together both C- and J-type
shock waves \citep{flower2003,giannini2006,gusdorf2008b}.
Predictions of H$_2$O line intensities in CJ-shocks are provided for a limited
set of conditions by \cite{gusdorf2011}: a qualitative comparison indicates that absolute
intensities and water abundances derived in these models are similar to those
observed in L1448. Clearly, the development of more specific models, suited for the shock regions
under investigation, would be needed for a better understanding of the water
properties highlighted by our observations. 

\subsection{Comparison with SiO emission}
\label{subsect:sio_emission}

Finally, we discuss the very different
line profile behaviour of H$_2$O and SiO at the two investigated positions 
(see Fig.~\ref{fig:h2o_sio_co} in Sect.~\ref{sect:line_profiles}).
At R4, the profile of the H$_2$O line at 557~GHz follows quite closely
the SiO profile. The SiO abundance is enhanced in shocks due to both sputtering of silicon from the
grain mantles, efficient at shock velocities as low as $\sim$~10 km~s$^{-1}$, and core grain disruption occurring
at higher shock speeds, of the order of $25-30$~km~s$^{-1}$ \citep[e.g.][]{jimenez2008,gusdorf2008a}.
The observed SiO emission in both the R4-LV and R4-HV components is therefore consistent with the range
of shock speeds we infer from the H$_2$O analysis.

At B2, the comparison between water and SiO shows a more complex situation, with a strong
variation of the SiO/H$_2$O ratio as a function of velocity. 
This behaviour probably reflects very different SiO/H$_2$O abundance ratios in the various velocity regimes, 
indicative of a different chemistry rather than a difference in the excitation conditions. 
In particular, the SiO emission at B2 is enhanced at the EHVs ($v \sim$~55~km~s$^{-1}$),
where no excess of water emission is detected. This is in contrast with the 
observations of water towards the driving source, where the EHVs water emission
is clearly detected (\citealt{kristensen2011}; Nisini et al. 2011, in prep.), which indicates
that EHV water emission is more localized than SiO and show up only in the regions with high water
column densities. It has been suggested that the EHV gas traces the primary 
proto-stellar jet, where molecules are rapidly synthesized from the
initial atomic gas \citep{tafalla2010}. 
Available chemical models for the chemistry in the primary jet, 
although simplified in the geometry and physical structure, predict a
low H$_2$O/SiO ratio for values of mass loss rates less than 10$^{-5}$~M$_{\odot}$~yr$^{-1}$ \citep{glassgold1991},
which is in agreement with the molecular mass flux rate of 10$^{-7}$~M$_{\odot}$~yr$^{-1}$, derived for L1448-mm
by \cite{dionatos2009}. 
Regarding the very low SiO associated with the wing emission, this is a bit puzzling given that these
wings extend up to 50~km~s$^{-1}$, i.e. where one would expect sputtering to be very efficient in
releasing silicon from grains. \cite{tafalla2010}, studying the chemistry of the L1448--R2 shock spot,
located in the red-shifted lobe symmetrically to the B2 position, point out a strong trend of increasing
SiO abundance with velocity, going from the lower velocities up to the EHV gas,
which is not followed by the other molecules they observed. 
It therefore appears that there are strong chemistry variations across the various velocity regimes at different 
positions that are not easily explained by current shock chemical models.

%
%______________________________________________________________

\section{Conclusions}

Herschel observations of several ortho- and para-H$_2$O transitions 
towards the outflow driven by the L1448 low-mass
proto-stellar system have been presented. 
These observations are part of the WISH key program.
Two shocked positions, R4 and B2, along the main L1448 outflow have been studied.
The main results of this work are the following:

\begin{enumerate}
\item The two investigated positions (R4 and B2) show physical and chemical differences. 
In particular, R4 exhibits strong variations in the excitation conditions
as a function of the LSR velocity and two velocity components can be distinguished in the water emission: 
a low-velocity component (R4-LV) and a high-velocity component (R4-HV).
The excitation at R4 decreases with velocity, 
i.e. the lines with higher upper level energies peak at lower LSR velocity, while the lines 
with lower upper level energies peak at higher LSR velocity.
\item The observed emission in both shocked positions is best represented by very dense 
($n_{\rm H_2} \sim 10^6-10^7$~cm$^{-3}$)
and warm ($T_{\rm kin}$=400--600~K) gas, having moderate H$_2$O column densities:
$N$(H$_2$O)$ \sim 10^{13}$~cm$^{-2}$ at R4-LV and $\sim 10^{14}-5$~10$^{14}$~cm$^{-2}$ 
at R4-HV and B2. These column density values correspond to 
H$_2$O/H$_2 \sim 0.5-1$~10$^{-5}$ at B2 and R4-HV, obtained from comparison with $N$(H$_2$)
derived from Spitzer observations \citep{giannini2011}.
\item The inferred physical conditions seem to be better reproduced by a J-type shock, where large 
compression factors leading to high density in the post-shock gas are expected. Moreover, the relatively 
low observed column densities and the PACS OH and [OI]~63$\mu$m observations 
are consistent with models where H$_2$O takes time to reform from fully dissociated gas.
However, more detailed shock models are needed. 
\item These {\it Herschel} H$_2$O observations provide evidence that water is unique in 
tracing very dense shock components, not traced by other molecules, such as CO and SiO, but rather 
traced by H$_2$.
In particular, H$_2$O shows strong differences with SiO in the excitation conditions and in the 
line profiles in the two observed shocked positions, pointing out to chemical variations 
across the various velocity regimes at different positions.
\end{enumerate}

\begin{acknowledgements}
The authors are grateful to the WISH internal referees Sylvain Bontemps and Bengt Larsson 
for their constructive comments on the manuscript and to Paola Caselli for useful discussions.
WISH activities in Osservatorio Astronomico di Roma are supported by the ASI project 01/005/11/0.
B.N. and G.S. also acknowledge financial contribution from the agreement
ASI-INAF I/009/10/0.

HIFI has been designed and built by a consortium of 
institutes and university departments from across Europe, Canada and the 
United States under the leadership of SRON Netherlands Institute for Space
Research, Groningen, The Netherlands and with major contributions from 
Germany, France and the US. Consortium members are: Canada: CSA, 
U.Waterloo; France: CESR, LAB, LERMA, IRAM; Germany: KOSMA, 
MPIfR, MPS; Ireland, NUI Maynooth; Italy: ASI, IFSI-INAF, Osservatorio 
Astrofisico di Arcetri- INAF; Netherlands: SRON, TUD; Poland: CAMK, CBK; 
Spain: Observatorio Astron{\'o}mico Nacional (IGN), Centro de Astrobiolog{\'i}a 
(CSIC-INTA). Sweden: Chalmers University of Technology - MC2, RSS $\&$ 
GARD; Onsala Space Observatory; Swedish National Space Board, Stockholm 
University - Stockholm Observatory; Switzerland: ETH Zurich, FHNW; USA: 
Caltech, JPL, NHSC.
\end{acknowledgements}

%\Online

\begin{appendix}

\section{Diagnostic diagrams}
\label{Asect:diagn_plots}

\begin{figure}[ht!]
   \centering
   \includegraphics[width=0.48\textwidth]{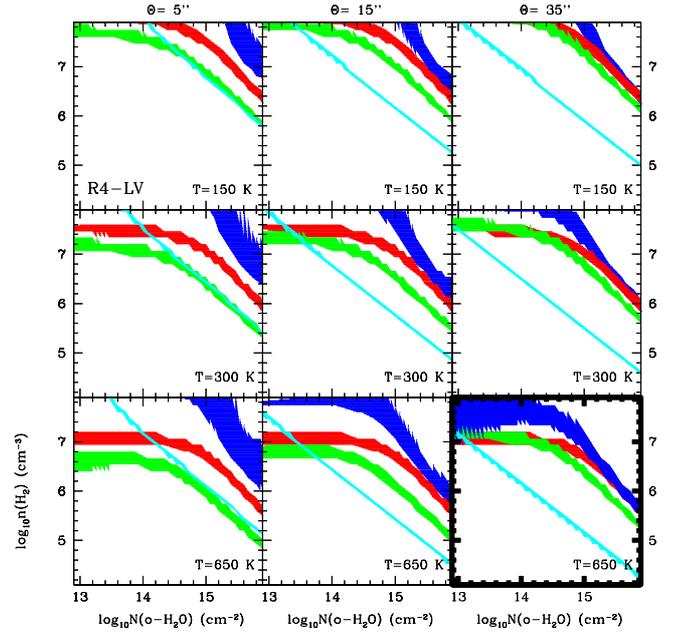}
   \caption{Diagnostic diagram to explore the parameter ($T_{\rm kin}$, $n_{\rm H_2}$, $N_{\rm ortho-H_2O}$, 
$\Theta$) space that fits the observed data of the R4-LV component.
The different panels are for three different values of the emission size 
($\Theta =$~5$^{\prime\prime}$, 15$^{\prime\prime}$ and 35$^{\prime\prime}$)
and three different temperatures ($T=$~150, 300 and 650~K).
Three line ratios are reported in each panel: 557GHz/1097GHz (green), 1670GHz/1097GHz (blue) and 1113GHz/1097GHz (red); 
and the integrated intensity of the 1097~GHz line (cyan). An error of 10\% is assumed for each line intensity.
The thick boxes in this diagram and in all the following diagrams mark 
the range of parameters for which all the line ratios and line intensity intersect, 
corresponding to the physical conditions that are consistent with the observed data.}
   \label{Afig:LV-diagnostic}
   \end{figure}

\begin{figure}[ht!]
   \centering
   \includegraphics[width=0.48\textwidth]{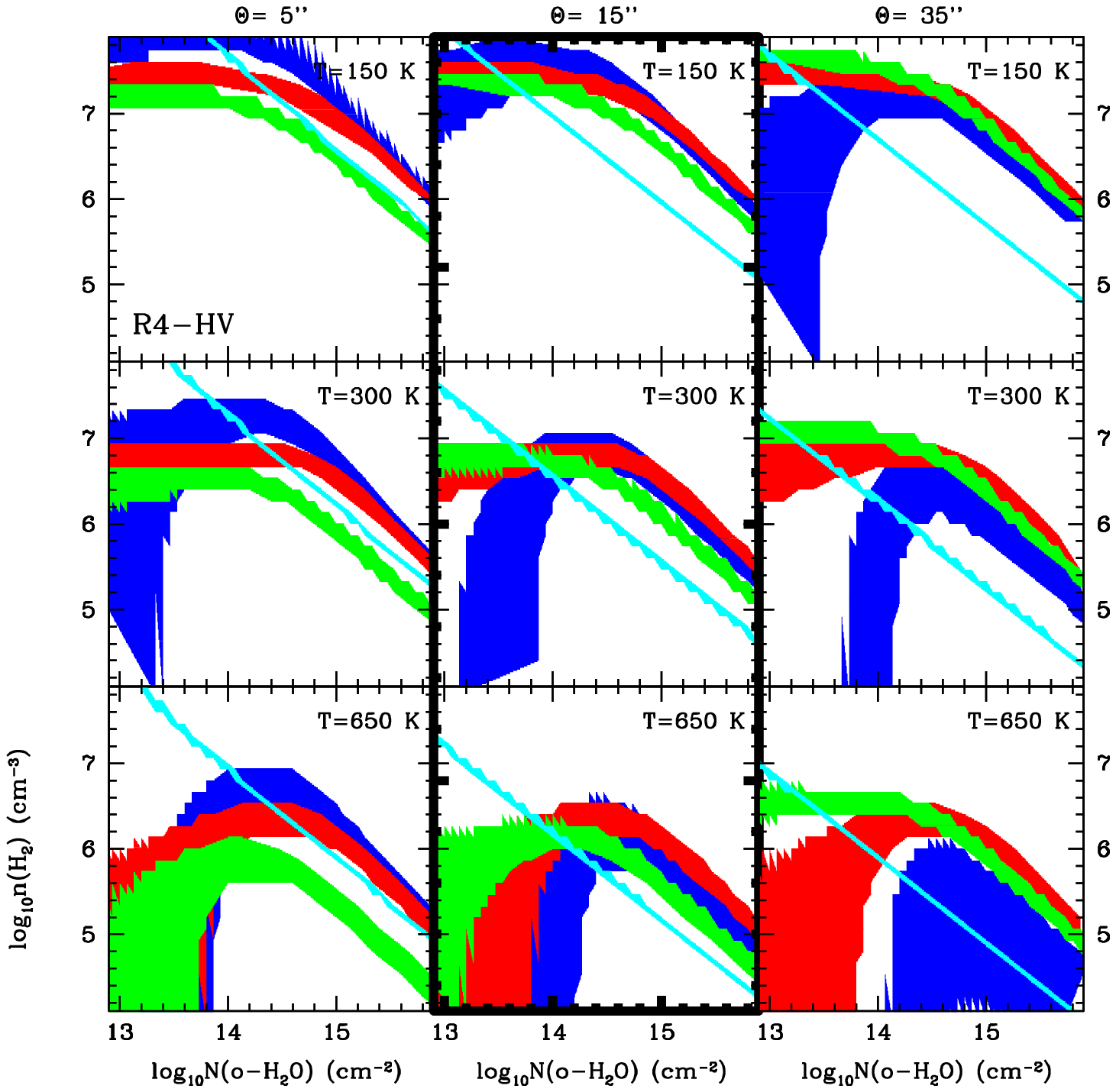}
   \caption{Same as Fig.~\ref{Afig:LV-diagnostic} for R4-HV.}
   \label{Afig:HV-diagnostic}
   \end{figure}

\begin{figure}[ht!]
   \centering
   \includegraphics[width=0.48\textwidth]{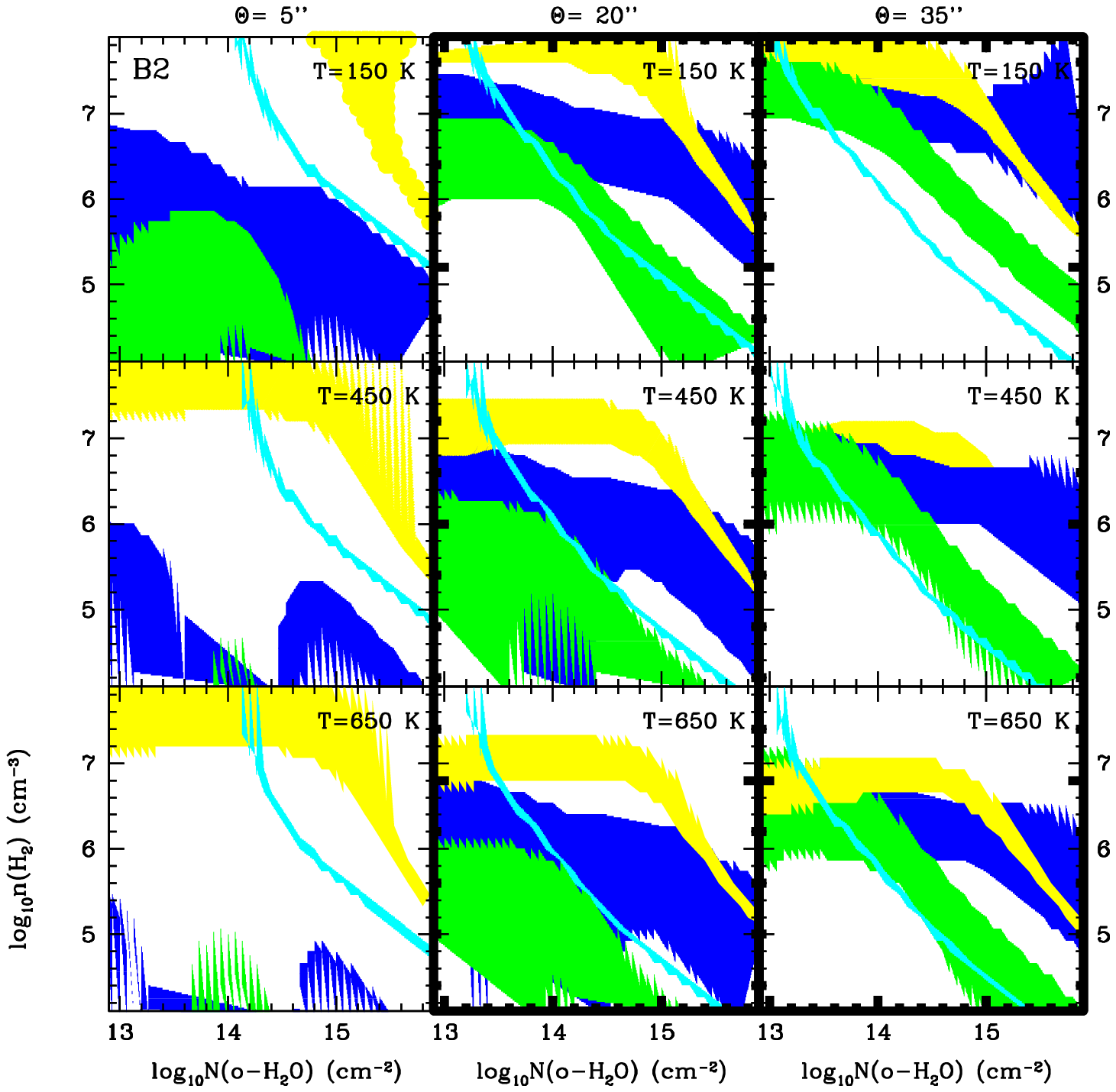}
   \caption{Diagnostic diagram to explore the parameter ($T_{\rm kin}$, $n_{\rm H_2}$, $N_{\rm ortho-H_2O}$, $\Theta$) space 
for the water emission at the B2 position.
The different panels are for three different values of the emission size ($\Theta =$~5$^{\prime\prime}$, 20$^{\prime\prime}$ and 35$^{\prime\prime}$)
and three different temperatures ($T=$~150, 450 and 650~K).
Three line ratios are reported in each panel: 557GHz/1113GHz (green),  1670GHz/1113GHz (blue) and 752GHz/1113GHz (yellow); 
and the integrated intensity of the 1113~GHz line (cyan). An error of 10\% is assumed for each line intensity.}
   \label{Afig:B2-diagnostic}
   \end{figure}

The results of the excitation analysis, presented in Sect.~\ref{sect:analysis}, 
are here illustrated through the use of diagnostic plots, employing both line ratios and 
integrated intensities of the water lines.
Each H$_2$O emission component, analyzed in the paper, is shown separately: 
the two velocity components in the water emission from the R4 position, R4-LV and R4-HV, 
are shown in Figs.~\ref{Afig:LV-diagnostic} and \ref{Afig:HV-diagnostic}, respectively, while 
the water emission from the B2 position is presented in Fig.~\ref{Afig:B2-diagnostic}.

   \begin{figure*}%[ht!]
   \centering
   \includegraphics[angle=-90, width=0.8\textwidth]{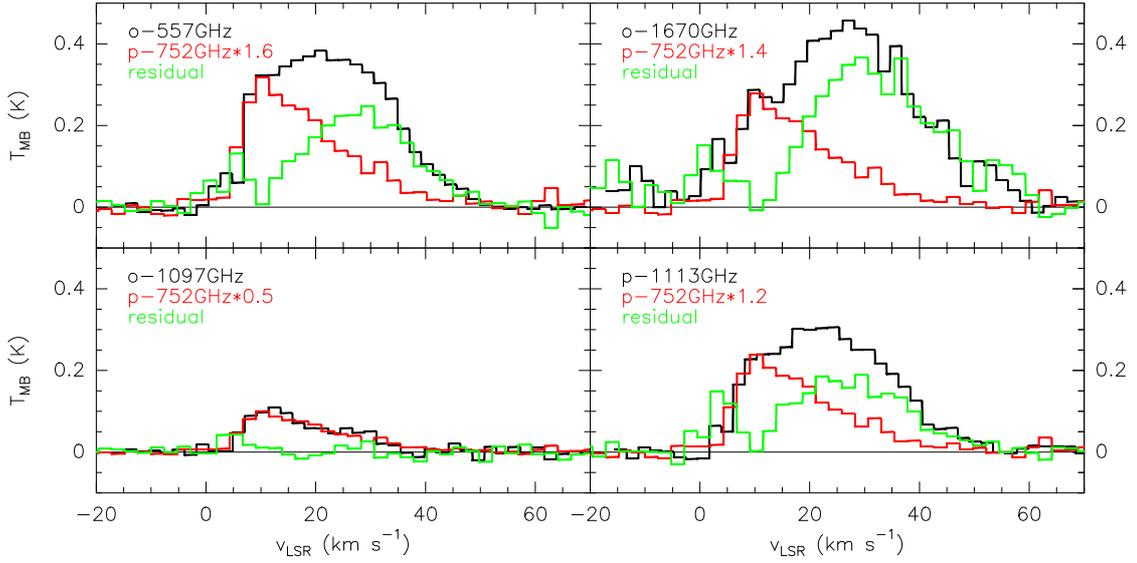}
   \caption{H$_2$O line profile (black) decomposition in a triangular low-velocity component (red), 
derived by fitting the 752~GHz water line to the low velocity emission profile, 
and a high-velocity component (green), obtained by subtracting the low-velocity component from the total water line profile.}
   \label{Afig:spectra_decomp}
   \end{figure*}
In particular, for each emission component we report a set of nine plots to  
explore the parameter ($T_{\rm kin}$, $n_{\rm H_2}$, $N_{\rm ortho-H_2O}$, $\Theta$) space 
and visualize the results given in Table~\ref{table:h2o_best_fit} and Fig.~\ref{fig:mod-obs}.
We considered the three most significant line ratios and one integrated line intensity.
In particular, for the R4-LV and R4-HV components, the 557GHz/1097GHz (green), 1670GHz/1097GHz (blue) and 
1113GHz/1097GHz (red) ratios and the integrated intensity of the 1097~GHz line are reported in each panel.
For the B2 component, the 557GHz/1113GHz (green), 1670GHz/1113GHz (blue) and 752GHz/1113GHz (yellow)
ratios and the integrated intensity of the 1113~GHz line are shown. 

At R4, to separate the contribution from the two velocity components with different excitation conditions 
along the line of sight (R4-LV and R4-HV), we divided the emission range of the water lines:
between 0 and 20 km~s$^{-1}$ for the low velocity component, R4-LV, and between 20 and 60 km~s$^{-1}$ for the 
high velocity component, R4-HV.

As an example, Fig.~\ref{Afig:LV-diagnostic} for R4-LV shows that, 
for a given temperature ($T$) and emission size ($\Theta$), 
at low H$_2$O column densities the line ratios only depend on the H$_2$ density, without any dependence on the H$_2$O
column density; while for higher H$_2$O column densities there is a degeneracy between the H$_2$O column 
density and the H$_2$ density, i.e. the line ratios depend on the product of both quantities. 
However, %the degeneragy is broken by 
different line ratios and the integrated intensity of one line 
%which also depends on both the gas density and the H$_2$O column density
allow to select only a limited region of the plane and therefore a limited range of physical conditions
(i.e. in the case of R4-LV, all the line ratios and the integrated intensity of the selected line 
only intersect in the bottom-right panel of Fig.~\ref{Afig:LV-diagnostic}).

The following conclusions can be drawn from the inspection of the diagrams in 
Figs.~\ref{Afig:LV-diagnostic}-\ref{Afig:B2-diagnostic}:
\begin{enumerate}
\item The water emission at R4-LV is consistent with an extended component, 
with a very dense gas (larger than 10$^7$ cm$^{-3}$) and low column density (a few 10$^{13}$ cm$^{-2}$).
The temperature of this extended R4-LV component is high, about 600~K.
\item R4-HV, at lower excitation than R4-LV, can be reproduced by a less extended emission
(about 15--20 arcsec) with a range of temperatures from 150~K to 650~K 
and either a higher density (larger than 10$^7$ cm$^{-3}$) and lower column density (less than 10$^{14}$ cm$^{-2}$)
or a lower density (around 10$^6$ cm$^{-3}$) and higher column density (a few 10$^{14}$ cm$^{-2}$), respectively. 
\item The emission at B2 appears to be consistent with an extended component (around 35 arcsec), 
with a wide range of temperatures, high density (larger than 10$^6$ cm$^{-3}$) and low column density 
(around or smaller than 10$^{14}$ cm$^{-2}$).
However, we know from the PACS map at 1670~GHz towards L1448 (see Fig.~\ref{fig:outflow}) that 
the water emission comes from a more compact region. 
Therefore, the solution for an emission size about 20 arcsec
was also considered, although it does not reproduce one of the line ratios (752GHz/1113GHz)
for any of the displayed temperatures. 
Thus to further constrain the physical parameters of the water emission at the B2 position,
some assumptions need to be made. 
As explained in Sect.~\ref{sect:analysis}, Spitzer mid-IR H$_2$ observations from \cite{giannini2011} 
has be used in this context to constrain the temperature to $\sim$ 450~K, assuming that 
water emission originates from the same gas emitting the mid-IR H$_2$ lines.
\end{enumerate}

   \begin{figure}
   \centering
   \includegraphics[width=0.48\textwidth]{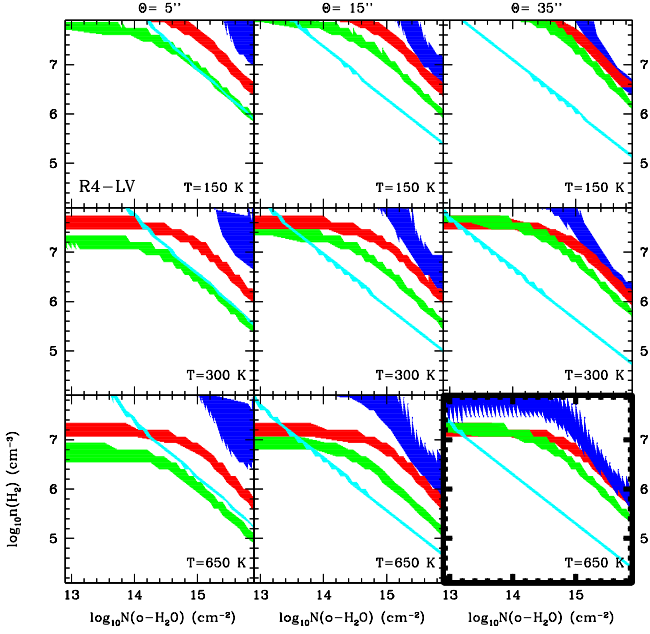}
   \caption{Diagnostic diagram to explore the parameter ($T_{\rm kin}$, $n_{\rm H_2}$, $N_{\rm ortho-H_2O}$, $\Theta$) space 
for the water emission at R4-LV, after dividing the two velocity components as explained in the text and 
illustrated in Fig.~\ref{Afig:spectra_decomp}.
The different panels are for three different values of the emission size ($\Theta =$~5$^{\prime\prime}$, 15$^{\prime\prime}$ and 35$^{\prime\prime}$)
and three different temperatures ($T=$~150, 300 and 650~K).
Three line ratios are reported in each panel: 557GHz/1097GHz (green),  1670GHz/1097GHz (blue) and 1113GHz/1097GHz (red); 
and the integrated intensity of the 1097~GHz line (cyan). An error of 10\% is assumed for each line intensity, to simplify the plots.}
   \label{Afig:LV-triang}
   \end{figure}
   \begin{figure}
   \centering
   \includegraphics[width=0.48\textwidth]{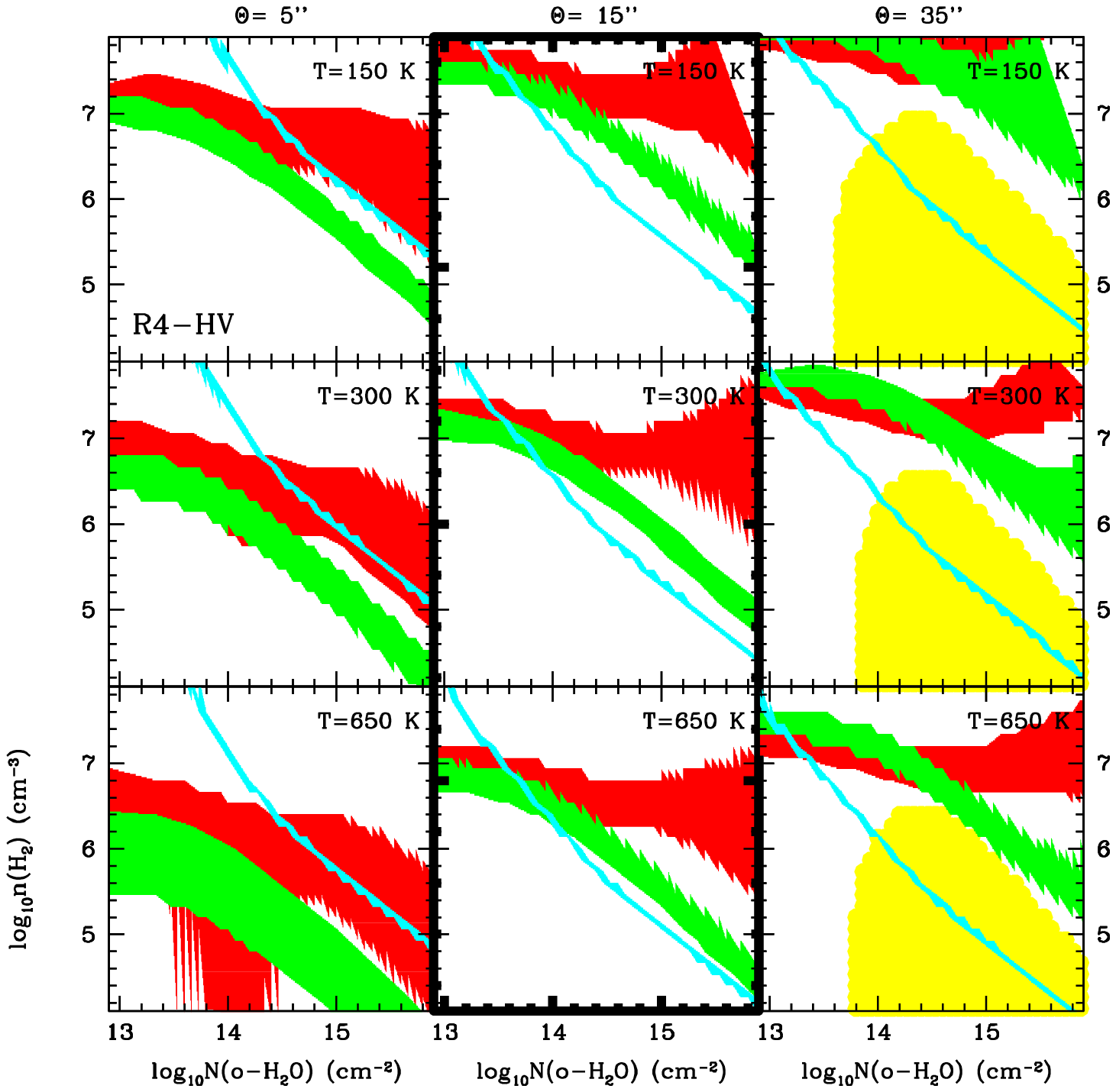}
   \caption{Same as Fig.~\ref{Afig:LV-triang} for R4-HV. Two line ratios are reported in each panel: 557GHz/1670GHz (green) and 
1113GHz/1670GHz (red); and the integrated intensity of the 1670~GHz line (cyan). Moreover the upper limits to the 
752GHz/1670GHz (yellow) line ratio is shown in the right panels, corresponding to $\Theta=$~35$^{\prime\prime}$.}
   \label{Afig:HV-triang}
   \end{figure}
Finally, as mentioned in Sect.~\ref{sect:analysis}, 
we also explored how the derived results can be
affected by the method used to separate the high and low velocity components in R4.
For that, we tried to deconvolve the line profiles into two different components 
based on the line shape: R4-LV being identified with the triangular profile 
seen especially in the higher excitation lines; 
while R4-HV corresponding to the high velocity emission component that is overlaid on the triangular 
shape in the lower excitation water lines.
Given the triangular line profile of the para-H$_2$O 2$_{11}-2_{02}$ line at 752~GHz, we 
assumed that all the emission in this line is coming from R4-LV. 
We thus used this line as a template for separating the R4-LV contribution in each water line profile, 
as illustrated in Fig.~\ref{Afig:spectra_decomp}:
we scaled the 752~GHz line to fit the low velocity triangular contribution 
(red curve) in each water line profile (black curve) and then subtracted it from the rest of 
the line, to obtain the R4-HV component (green curve).
In this way we decomposed every water line profile in the two contributions from the R4-LV and R4-HV gas components.
For each water line, both derived spectra have thus been integrated over the whole emission range to 
derive the integrated intensities of the two components.

We thus compared the derived line ratios and one integrated line intensity of R4-LV (Fig.~\ref{Afig:LV-triang}) 
and R4-HV (Fig.~\ref{Afig:HV-triang}) with the models.
The considered integrated line intensities for each component are the 1097~GHz line for the R4-LV and the 1670~GHz line for R4-HV. 
Moreover, for R4-HV, we show also the upper limits to the 
752GHz/1670GHz line ratio (yellow), 
derived from the 3$\sigma$ upper limit of the 
752GHz integrated intensity, in the only three panels where it is significant and can be used to 
discriminate between the different models.

The plots clearly highlight the same results obtained in the previous section, 
meaning that the approach we use to separate the two velocity components at 
different excitation at the R4 position does not affect the results of the analysis.
This is mostly because the two velocity components are well separated in velocity 
(R4-LV peaks at about 10~km~s$^{-1}$ and R4-HV at about 25~km~s$^{-1}$)
and therefore each of them dominates the water emission in the relative velocity range 
(0--20~km~s$^{-1}$ for R4-LV and 20--60~km~s$^{-1}$ for R4-HV) we chose in Sect.~\ref{sect:analysis}.

\end{appendix}

\end{document}